# Momentum Resolved Superconducting Energy Gaps of Sr$_2$RuO$_4$ from Quasiparticle Interference Imaging

Rahul Sharma[1,2]†, Stephen D. Edkins[3]†, Zhenyu Wang[4]†, Andrey Kostin[1,2], Chanchal Sow[5,6], Yoshiteru Maeno[5], Andrew P. Mackenzie[7], J.C. Séamus Davis[8,9] and Vidya Madhavan[4]

1. LASSP, Department of Physics, Cornell University, Ithaca, NY 14853, USA.
2. CMPMS Department, Brookhaven National Lab., Upton NY, USA
3. Department of Applied Physics, Stanford University, Stanford, CA 94305, USA
4. Department of Physics, University of Illinois, Urbana, Illinois 61801, USA
5. Department of Physics, Kyoto University, Kyoto 606-8502, Japan
6. Department of Physics, IIT-Kanpur, Uttar Pradesh, 208016 India
7. Max Planck Institute for Chemical Physics of Solids, D-01187 Dresden, Germany
8. Clarendon Laboratory, University of Oxford, Parks Rd., Oxford, OX1 3PU, UK
9. Department of Physics, University College Cork, Western Road, Cork T12R5C, Ireland.

**Sr$_2$RuO$_4$ has long been the focus of intense research interest because of conjectures that it is a correlated topological superconductor. It is the momentum space ($k$-space) structure of the superconducting energy gap $\Delta_i(k)$ on each band *i* that encodes its unknown superconducting order-parameter. But, because the energy scales are so low, it has never been possible to directly measure the $\Delta_i(k)$ of Sr$_2$RuO$_4$. Here we implement Bogoliubov quasiparticle interference (BQPI) imaging, a technique capable of high-precision measurement of multiband $\Delta_i(k)$. At T=90 mK we visualize a set of Bogoliubov scattering interference wavevectors $q_j: j = 1 - 5$ consistent with eight gap nodes/minima, that are all closely aligned to the $(\pm 1, \pm 1)$ crystal-lattice directions on both the *α*- and *β*-bands. Taking these observations in combination with other very recent advances in directional thermal conductivity (E. Hassinger *et al.* Phys. Rev. X 7, 011032 (2017)), temperature dependent Knight shift (A. Pustogow *et al.* Nature 574, 72 (2019)), time-reversal symmetry conservation (S. Kashiwaya *et al.* Phys. Rev B, 100, 094530 (2019)) and theory (A.T. Romer *et al.* Phys. Rev. Lett. 123, 247001 (2019); H. S. Roising *et al.* Phys. Rev. Research 1, 033108 (2019), O. Gingras *et al.* Phys. Rev. Lett. 123, 217005 (2019)), the BQPI signature of Sr$_2$RuO$_4$ appears most consistent with $\Delta_i(k)$ having $d_{x^2-y^2}$ ($B_{1g}$) symmetry.**

*Keywords:* Strontium Ruthenate, Quasiparticle Interference, Superconducting Energy Gaps

**Significance Statement:**

*$Sr_2RuO_4$ has been widely studied as a candidate correlated topological superconductor. However, the momentum space structure of the superconducting energy gaps which encode both the pairing mechanism and its topological nature, have proven impossible to determine by conventional techniques. To address this challenge, we introduce Bogoliubov quasiparticle scattering interference visualization at millikelvin temperatures. We discover that the $\alpha$- and $\beta$-bands of $Sr_2RuO_4$ support thermodynamically prevalent superconducting energy gaps, and that they each contain four gap nodes (or profound minima) that are contiguous to the $(0,0) \to (\pm 1, \pm 1)\pi/a$ lines in momentum space. In the context of other recent advances, these observations appear most consistent with a $d_{x^2-y^2}$ order-parameter symmetry for $Sr_2RuO_4$.*

**1**     Determining the structure and symmetry of the superconducting energy gaps $\Delta_i(k)$ for $Sr_2RuO_4$ has been a longstanding objective[1-4], but one upon which radically new perspectives have emerged recently. The linearity with temperature of electronic specific heat capacity at lowest temperatures[5], the temperature dependence of London penetration depth[6], the attenuation rate of ultrasound[7] and field-oriented specific heat measurements[8] have long implied the existence of nodes (or profound minima) somewhere in $\Delta_i(k)$. But recent thermal conductivity measurements further indicate that these nodes/minima are oriented parallel to the crystal c-axis[9]. Moreover, in-plane [17]O nuclear magnetic resonance reveals a very substantial drop of the Knight shift[10] below $T_c$. And no cusp occurs in the superconducting critical temperature under uniaxial strain[11,12]. Finally, current-field inversion experiments using Josephson tunnel junctions indicate that time reversal symmetry (TRS) is preserved[13]. This phenomenology is in sharp contradistinction to the $Sr_2RuO_4$ *ancien regime*, under which [17]O Knight shift[14] and spin-polarized neutron scattering[15] reported no diminution in spin susceptibility below $T_c$, and where muon spin rotation[16] and Kerr effect[17] indicated TRS breaking. Therefore, an extensive reassessment of the theory of Sr$_2$RuO$_4$ superconductivity has quickly materialized[18-23].

**2**     Although the crystal is isostructural with the *d*-wave high temperature superconductor La$_2$CuO$_4$ (Fig. 1A), for Sr$_2$RuO$_4$ the Fermi surface (FS) consists of three sheets[24,25,] (Fig. 1b). Hybridization between the two quasi-one-dimensional (1D) bands that originate from the Ru $d_{xz}$ and $d_{yz}$ orbitals, leads to the electron-like β-band surrounding the Γ-point (red) and hole-like α-band surrounding the X point (blue); similarly, the Ru $d_{xy}$ orbitals generate the electron-like, quasi-two-dimensional (2D) γ-band surrounding the Γ-point (green). Correctly representing the electron-electron interactions is then a complex challenge. On-site and inter-site Coulomb interactions are pervasive, Hund's coupling between the Ru *d*-orbitals generates orbital selective phenomena rendering the γ-band significantly more correlated than the *α:β*-bands[26,27], and spin-orbit coupling plays a significant role throughout[26]. Contemporary theories[18,19,20,28,29] consider various combinations of these interactions to achieve their $\Delta_i(k)$ predictions, focusing on the dependence of symmetry of the predominant $\Delta_i(k)$ on the interplay between them. Weak-coupling analyses[28,29] of Hamiltonians parameterized by the ratio *ρ=J/U* (*U* and *J* are the on-site Coulomb and Hund's interaction energies) find that the preferred order parameters exhibit $E_u$ (chiral $p-$wave) symmetry with $B_{1g}$ ($d_{x^2-y^2}$) symmetry as a subdominant solution[28]; and $E_u$ (chiral $p-$wave) or $A_{1u}$ (helical $p-$wave) symmetry but with $B_{1g}$ ($d_{x^2-y^2}$) symmetry also as a subdominant solution[29]. More recent theories parameterized by both *ρ=J/U* and spin-orbit coupling λ, find that the order parameters filling large (but different) portions of the ρ–λ phase space exhibit $B_{1g}$ ($d_{x^2-y^2}$) symmetry[18,19,20] , and $A_{1u}$ (helical $p-$wave) symmetry[18,19] or even more complex spin-triplet orders[20]. One surprising consequence is that the field-in-plane Knight shift does not discriminate strongly between $\Delta_i(k)$ having $d_{x^2-y^2}$ (even $-$ parity) or $p_{helical}$ (odd $-$ parity) order parameters[18,19]. Obviously, what could discriminate between all these different order parameter symmetries is the fully detailed structure of $\Delta_i(k)$, as shown e.g. in Figure 2 of Ref. 18 or Figure S3 of Ref. 19.

*3*    However, although critical to testing advanced theories[18-23] for superconductivity in Sr2RuO4, the $\boldsymbol{k}$-space structure of $\Delta_\alpha(\boldsymbol{k}); \Delta_\beta(\boldsymbol{k}); \Delta_\gamma(\boldsymbol{k})$ has never been measured directly. Basically, this is because the maximum magnitude of any of these gaps[30,31] is $|\Delta| \leq 350\ \mu eV$ so that temperature $T \lesssim 100\ mK$ and energy resolution with $\delta E \lesssim 100\ \mu eV$ are required to spectroscopically detect strongly anisotropic $\boldsymbol{k}$-space gap structures and/or their gap minima. Thus, techniques capable of band-resolved, high resolution superconducting $\Delta(\boldsymbol{k})$ determination, and specifically of distinguishing the orientation of any gap minima on different bands, are required. Bogoliubov quasiparticle interference imaging[32-38] has been proposed[39,40,41] to achieve these objectives for Sr2RuO4, as it has the proven capability of measuring extremely anisotropic[33-38], multiband[35,36,38] superconducting energy gaps with energy resolution[36,38] $\delta E \lesssim 75\ \mu eV$. Intuitively, this is possible because, when a highly anisotropic $\Delta_k$ opens on a given band, Bogoliubov quasiparticles $|\boldsymbol{k}(E)\rangle$ exist in the energy range $\Delta_k^{min}<E<\Delta_k^{max}$. Within this range, interference of impurity-scattered quasiparticles produces characteristic real space ($\boldsymbol{r}$-space) modulations in the density of electronic states[32,39,40,41] $\delta N(\boldsymbol{r}, E)$. The Bogoliubov quasiparticle dispersion $E_i(k)$ then exhibits closed constant-energy-contours (CEC) surrounding Fermi surface $\boldsymbol{k}$-points where minima in $\Delta_k$ occur. These $\boldsymbol{k}$-space locations can be determined because $\delta N(\boldsymbol{r}, E)$ modulations occur at the set of wavevectors $q_j(E)$ connecting them. These $q_j(E)$ are identified from maxima in $\delta N(\boldsymbol{q}, E)$, the power spectral density Fourier transform of $\delta N(\boldsymbol{r}, E)$.

*4*    For Sr2RuO4, BQPI signatures of different types of gap structures, for example $\Delta_\alpha(\boldsymbol{k}); \Delta_\beta(\boldsymbol{k})$, may be anticipated by using a pedagogical Hamiltonian $H(k) =$

$\sum_k \psi^\dagger(k)\hat{H}(k)\psi(k)$ where

$$\hat{H}(k) = \begin{pmatrix} \epsilon_\alpha(k) & \Delta_\alpha(k) & 0 & 0 \\ \Delta_\alpha^*(k) & -\epsilon_\alpha & 0 & 0 \\ 0 & 0 & \epsilon_\beta(k) & \Delta_\beta(k) \\ 0 & 0 & \Delta_\beta^*(k) & -\epsilon_\beta(k) \end{pmatrix} \quad (1)$$

and $\epsilon_\alpha(k)$, $\epsilon_\beta(k)$ are the band dispersion for the α:β-bands[39,40,41]. The unperturbed Green's function is $G^0(k,\epsilon) = [(\epsilon + i\delta)I - \hat{H}(k)]^{-1}$ where I is identity matrix and δ is the energy width broadening parameter. Both interband and intraband scattering could be considered using a T-matrix for all scattering processes as:

$$T^{-1}(\omega) = I \otimes (V_{intra}I + V_{inter}\sigma_x)^{-1} - \int \frac{dk}{2\pi} G^0(k,\omega) \quad (2)$$

But interband scattering between the α:β and γ bands has not been the subject of any theoretical analysis for Sr$_2$RuO$_4$ (Refs 39,40,41) hence we do not consider it here. The Fourier transform of $\delta N(r, E)$ modulations caused by scattering interference of Bogoliubons can be predicted from Eqns. 1 and 2 as:

$$\delta N(q, E) = -Im\left[Tr\left(\int \frac{dk}{2\pi} G^0(k+q,\omega)T(\omega)G^0(k,\omega)\right)\right] \quad (3)$$

(SI Section I). For example, Fig. 1c represents BQPI for an anisotropic energy gap $\Delta_\gamma(k)$ on the γ-band, while Fig. 1d represents a BQPI model with anisotropic energy gaps $\Delta_\alpha(k); \Delta_\beta(k)$ on the α:β-bands. The experimental challenge is to visualize Bogoliubov scattering interference in Sr$_2$RuO$_4$ and, through comparison with $\delta N(q, E)$ predictions[39,40,41], to determine $\Delta_i(k)$.

**5**     To do so, we insert high quality, single crystals of Sr$_2$RuO$_4$ (T$_c$ = 1.45K) into a dilution-refrigerator-based spectroscopic imaging scanning tunneling microscope (SI-STM), and cleave them in cryogenic ultra-high vacuum at $T \lesssim 1.8\ K$. This typically reveals an atomically flat SrO cleave surface (Fig. 1a) although sometimes the RuO$_2$ termination layer occurs[31]. At the SrO termination surfaces used throughout these studies (e.g. Fig. 1a), the tip-sample differential tunneling conductance $g(r, E) \equiv dI/dV(r, E = eV)$ is imaged to visualize scattering interference induced modulations $g(r, E) \propto \delta N(r, E)$. In the normal state, $g(r, E)$ measurements in the range $-20meV < E < 20meV$ reveal $g(q, E) \propto \delta N(q, E)$ (Fig. 2a) with predominant scattering wavevectors $\boldsymbol{q}(E)$ shown as red and blue arrows. Quantitative comparison to the known FS $\boldsymbol{k}(E=0)$ wavevectors[25], reveals that these arise from intraband scattering in both the β-band and the α-band (Fig. 1b) (SI Section II). As in previous QPI studies of normal-state Sr$_2$RuO$_4$, the γ-band is virtually undetectable, probably because the $d_{xy}$ character leads to small wavefunction overlap for tunneling into the STM tip[31]. In any case, the α:β-bands are directly identifiable from their normal state scattering interference wavevectors, throughout all the BQPI studies reported below.

**6**     To measure $\Delta_i(\boldsymbol{k})$, we cool each sample to T=90 mK (SI section III) and typically measure $g(\boldsymbol{r}, E) \propto \delta N(\boldsymbol{r}, E)$ on a 128x128 grid in a 20nm field of view. Typical junction formation parameters for these $g(\boldsymbol{r}, E)$ measurements are $I_S = 40pA; V_S = 1mV$, and $|E| = 0, 100\ \mu eV, 200\ \mu eV, 300\ \mu eV, 400\ \mu eV$ spanning the maximum superconducting energy gap (SI Section III). The actual electron temperature is manifestly well below ~100μeV/3.5k$_B$ or ~300mK because these BQPI images are distinct when the DC bias is changed in energy steps of 100 $\mu eV$. A representative point spectrum from such a map is shown in Fig. 2c, showing the typical[30,31] energy gap maximum $\Delta_{max} \approx 350\ \mu eV$. Figure 2d shows a typical measured $g(\boldsymbol{q}, E = 100\mu eV)$ deep within this superconducting gap. It is highly distinct from the $g(\boldsymbol{q}, E)$ measured near E$_F$ in the normal state (e.g. Fig. 2a) or at $E \gg 350\ \mu eV$ in the superconducting state (Fig. 4e), with many robust new $\boldsymbol{q}$-space features. Differences in signal intensity between $g(\boldsymbol{q}, E = 400\mu eV)$ measured in the normal and superconducting states occur due to the greatly reduced bias modulation amplitude required for the latter. Most importantly, the distinct $g(\boldsymbol{q}, E)$ at $|E| = 0, 100\ \mu eV, 200\ \mu eV, 300\ \mu eV$ at T=90mK hold the key to understanding the energy gap structure of Sr$_2$RuO$_4$ using Bogoliubov scattering interference[39,40,41]. At the most elementary level, Fig. 2d reveals spectroscopically that, consistent with a wide variety of other techniques[5,7,8,9], a strong Bogoliubov quasiparticle density of states exists deep within the superconducting gap of this material.

**7**     To aid with interpretation of these $g(\boldsymbol{q}, E)$ data, we explore a pedagogical model for $\Delta(\boldsymbol{k})$ having gap zeros along $(\pm 1, \pm 1)$ on $\alpha$:$\beta$-bands (Fig. 3a). In Fig. 3a the hypothetical gap magnitudes $|\Delta_\alpha(\boldsymbol{k})|, |\Delta_\beta(\boldsymbol{k})|$ are indicated by the thickness of the curves overlaid on the $\alpha$:$\beta$ FS. Figure 3b identifies the consequent $\boldsymbol{k}$-space regions where, because of minima in $\Delta_\alpha(\boldsymbol{k})$ and $\Delta_\beta(\boldsymbol{k})$, significant quasiparticle density of states is expected as $E \to 0$. The key BQPI wavevectors $q_j; j = 1,2..5$ (Fig. 3b) then connect these $\boldsymbol{k}$-space locations as shown. Figure 3c shows typical evaluations of $\delta N(\boldsymbol{q}, E)$ from Eqn. 3 for this model, with the key BQPI wavevectors overlaid. Here $q_1, q_2, q_3$ (Fig. 3b) occur due to the gap minima/nodes on the $\beta$-band, while $q_4, q_5$ (Fig. 3b) occur due to gap minima/nodes on the $\alpha$-band. Observation of BQPI intensity in $g(\boldsymbol{q}, E)$ data at these specific wavevectors $q_j; j = 1,2..5$ would give direct evidence for a superconducting energy gap structure (Fig. 3a) with gap minima/nodes along the $(\pm 1, \pm 1)$ on the $\alpha$:$\beta$ -bands of Sr$_2$RuO$_4$.

**8**     Figures 4a-h contain the key experimental results of this study: the measured $g(\boldsymbol{q}, E)$ at multiple energies within the superconducting gap of Sr$_2$RuO$_4$, at T=90 mK. The $g(\boldsymbol{q}, E = 1\,meV)$ in Fig. 4a is shown for comparison. Predictions from Eqn. 3 for $\delta N(\boldsymbol{q}, E)$ with the gap model in Fig. 3a are shown at corresponding energies to the measured $g(\boldsymbol{q}, E)$, in Figures 4e-h. The simultaneously measured $g(\boldsymbol{q}, E = 1meV)$ exhibits direct signatures of α:β-band scattering interference, as identified from our normal state studies (SI Section II). Since the electron tunneling manifestly occurs to the $\alpha$:$\beta$-bands and simultaneously exhibits a single-particle spectrum showing gap maximum $\Delta_{max} \approx 350\,\mu eV$ (Fig. 2c), we conclude that this superconducting gap is hosted by the $\alpha$:$\beta$-bands[31]. And, because $\Delta_{max} \approx 350\mu eV$ is a consistent gap maximum for the bulk superconducting critical temperature T$_c$=1.45K (because $2\Delta_{max}/kT_c \approx 4$), this indicates that $\Delta_\alpha(\boldsymbol{k}); \Delta_\beta(\boldsymbol{k})$ are principal energy gaps of Sr$_2$RuO$_4$.

**9**     Then, when Bogoliubov scattering interference is visualized at subgap energies $|E| < \Delta_{max}$, a new and distinctive $g(\boldsymbol{q}, E)$ pattern emerges. It exhibits clear maxima at specific $\boldsymbol{q}$-vectors (Fig. 4b,c,d) that evolve but do not disappear as $E \to 0$.

Theories of Sr$_2$RuO$_4$ BQPI demonstrate how these $q$-vectors encode the direction of the gap minima in $\Delta_\alpha(k)$; $\Delta_\beta(k)$, and also predict a very weak dispersion of the sub-gap $g(q, E)$ with energy[39,40,41]. The observed pattern of $g(q, E = 100 \mu eV)$ maxima in Fig. 4d is quite representative, and conforms to predicted $\delta N(q, E = 100 \mu eV)$ of the energy gap model in Fig 3. Specifically, in Fig. 5a the predicted BQPI wavevectors $q_1, q_2, q_3, q_4$ and $q_5$ from the $\alpha$:$\beta$-band model with nodes/minima along $(\pm 1, \pm 1)$ (circles), are compared to the locations of five distinct local maxima in $g(q, E = 100 \mu eV)$ in Fig. 5b and found to be in good agreement. The immediate implication is that eight nodes/minima occur in $\Delta_\alpha(k)$; $\Delta_\beta(k)$ at the locations where the $\alpha$:$\beta$-bands cross the $(0,0) \to (\pm 1, \pm 1)\pi/a$ symmetry axes. Because the measured $g(q, E)$ are distinct for $E = 0, 100 \, \mu eV, 200 \, \mu eV, 300 \, \mu eV$ (Fig. 4), the energy resolution $\delta E$ is demonstrably $\delta E < 100 \, \mu eV$, while from the measurement parameters we estimate that $\delta E \lesssim 75 \, \mu eV$. This means that if minima (as opposed to nodes) occur in $\Delta_\alpha(k)$ and $\Delta_\beta(k)$, they exist below the energy scale |E|=75 μeV. Moreover, analysis of the $g(q, E = 0)$ data shown in Fig. 5c indicates that all eight gap minima/nodes have an angular displacement about (0,0) in $k$-space, within approximately $\pm 0.05$ *rad* from the $(0,0) \to (\pm 1, \pm 1)\pi/a$ lines (SI Section IV). No features expected of $\Delta_\gamma(k)$ (SI Section II) are detected. As to the signature in $g(q, 0)$ of the predicted minima on $\Delta_\beta(k)$ in an odd-parity state (see Figure 2 of Ref. 18, Figure S3 of Ref. 19, Figure 5 of Ref. 31), these are expected to appear as $g(q, 0)$ maxima at wavevectors at least $\pm 0.1$ *rad* away from the $(0,0) \to (\pm 1, \pm 1)\pi/a$ lines[18,19,28,31]; or if the energy resolution is insufficient to resolve them, they should exhibit as a broad arc connecting these $g(q, 0)$ maxima. As discussed in SI Section V, neither of these signatures has been detected within the available signal to noise ratio. Moreover, in the same models[18,19,28,31] the minimum which occurs on $\Delta_\alpha(k)$ is typically shallow, whereas the measured minimum on $\Delta_\alpha(k)$ is deep reaching to within 75 μeV of zero (Fig. 5c). Therefore, a gap structure for both $\Delta_\alpha(k)$ and $\Delta_\beta(k)$ as shown in Fig. 5d, appears most consistent with our present data.

**_10_**     In this project, we provide the first momentum-resolved spectroscopic measurements of the superconducting gap structure in Sr$_2$RuO$_4$. They reveal eight nodes or deep minima in $\Delta_\alpha(\mathbf{k})$ and $\Delta_\beta(\mathbf{k})$ which occur in close proximity to where the $\alpha$:$\beta-$ bands cross the $(0,0) \to (\pm 1, \pm 1)\pi/a$ lines. In light of recent thermal conductivity[9], Knight shift[10], current-field reversal[13] experiments, and advanced theory[18,19,20,28,29], several key implications emerge from this observation. If time-reversal symmetry were actually broken[14,15,16,17] by $\Delta_i(\mathbf{k})$ of Sr$_2$RuO$_4$ but the order-parameter has even parity[10], then $s' + id_{x^2-y^2}$ (Ref. 18) or $d_{xz} + id_{yz}$ (Ref. 42) states would be plausible. Based on our BQPI data along with thermodynamic/transport studies[6,7,8,9], $d_{xz} + id_{yz}$ appears inconsistent because of its circumferential nodes in the $k_x$:$k_y$ plane, but $s' + id_{x^2-y^2}$ (Ref. 18) might be consistent. However, for such order parameters the transition temperature should split under a crystal-symmetry-breaking field, but that effect is reportedly absent in multiple relevant studies[11,12,43,44,45,46]. On the other hand, if time-reversal symmetry is preserved[13], the BQPI data (Figs 3,4) are most consistent with a helical odd-parity $p_{helical}$ order-parameter[18,19,28] with $A_{1u}$ symmetry, or an even-parity $d_{x^2-y^2}$ order-parameter[18,19,20,28] with $B_{1g}$ symmetry. In terms of the detailed $\mathbf{k}$-space structure of $\Delta_i(\mathbf{k})$ these two cases are distinct. The former exhibits minima but not nodes on the $\alpha$: $\beta-$ bands, their $\mathbf{k}$-space locations are not constrained by crystal symmetry, and the minima on different bands are not necessarily co-aligned in $\mathbf{k}$-space[18,19]. The latter exhibits true nodes on both the $\alpha-$ and $\beta-$ bands, whose $\mathbf{k}$-space locations are constrained precisely by crystal symmetry to lie along the $(\pm 1, \pm 1)$ directions. Our BQPI data (Fig. 4) implies that the four energy-gap minima/nodes of both $\Delta_\alpha(\mathbf{k})$ and $\Delta_\beta(\mathbf{k})$ exist below the energy scale |E|=75µeV, and that they occur within an angular distance from the $(0,0) \to (\pm 1, \pm 1)\pi/a$ $\mathbf{k}$-space lines of approximately $\pm 0.05$ *rad.* Overall, therefore, these observations appear most consistent with a $d_{x^2-y^2}$ order-parameter symmetry for Sr$_2$RuO$_4$.

**FIGURE CAPTIONS**

**Figure 1 Electronic structure of superconducting Sr$_2$RuO$_4$**

a. Topographic image of surface of Sr$_2$RuO$_4$ recorded at V$_s$ = 100mV and I$_s$ = 100pA showing SrO plane and defects which are Sr vacancies. All experiments reported in this paper are carried out under equivalent topographic conditions.
b. Model Fermi surface of Sr$_2$RuO$_4$ showing α (blue), β (red) and γ (green) bands.
c. Pedagogical model of a superconducting energy gap on γ-band with gap minima along ($\pm$1,0); (0,$\pm$1).
d. Pedagogical model of superconducting energy gaps on α:β-bands with gap minima along ($\pm$1,$\pm$1).

**Figure 2 Visualizing QPI from α:β bands and in the superconducting state**

a. Measured $g(\boldsymbol{q}, E)$ at T=2.1 K and E=20meV in normal state. Arrows show the features resulting from quasiparticle scattering from α (blue) and β (red) band.
b. Fermi surface showing α:β bands in red and blue respectively. Major scattering vectors as detected in experiments are overlaid.
c. Spatially averaged superconducting tunneling spectrum showing the full energy gap Δ ≈ 350 µ$eV$ measured at T=90mK.
d. Measured $g(\boldsymbol{q}, E)$ at T=90mK and $E = 100\mu eV$, deep within the superconducting energy-gap revealing the highly distinct Bogoliubov quasiparticle interference pattern of Sr$_2$RuO$_4$.

**Figure 3 Pedagogical Bogoliubov Scattering Interference Model**

a. Gap magnitude on the Fermi surface for α:β band with gap minima along ($\pm$1,$\pm$1).
b. Regions of significant quasiparticle density $E \rightarrow 0$ for *α:β*-bands when gapped as shown in a. Major scattering vectors are labeled as $q_j$ ; $j = 1,2,3,4,5$
**c.** Calculated $g(\boldsymbol{q}, E)$ pattern from Eqn. 3 for α:β band from the model in *a* at $E = 100\mu eV$. Key scattering wavevectors are indicated by $q_j$ ; $j = 1,2,3,4,5$.

**Figure 4 Imaging Bogoliubov Scattering Interference of Sr$_2$RuO$_4$**

a. to d. Measured $g(\boldsymbol{q}, E)$ images at T=90mK in superconducting state of Sr$_2$RuO$_4$ at E = 1meV, 300μeV, 200μeV and 100μeV. Red crosses denote Bragg peaks. Typically, the features at lowest $|\boldsymbol{q}|$ in experimental $g(\boldsymbol{q}, E)$ represent long range disorder/drift in the real space rather than any specific low $|\boldsymbol{q}|$ scattering interference. Moreover, the overall signal intensity here is weak because: (a) the density of impurity atoms necessary to avoid suppression of T$_c$ is very low and, (b) the low bias voltages and modulations required to visualize BQPI at these extremely low energy scales and with high energy resolution result in greatly increased averaging times per dI/dV spectrum.

e. to h. Predicted $\delta N(\boldsymbol{q}, E)$ for α:β-bands with minima along $(\pm 1, \pm 1)$ at E = 1meV, 300μeV, 200μeV and 100μeV. Red crosses denote RuO$_2$ Bragg peaks.

**Figure 5 Predominant $\Delta_\alpha(\boldsymbol{k}), \Delta_\beta(\boldsymbol{k})$ with Gap Minima/Nodes along $(\pm 1, \pm 1)$**

a. Predicted $\delta N(\boldsymbol{q}, E)$ for $\Delta_\alpha(\boldsymbol{k}), \Delta_\beta(\boldsymbol{k})$ at E = 100μeV with red (blue) circles denoting the features arising from scattering arising from *α:β*-bands.

b. Measured $g(\boldsymbol{q}, E)$ pattern at E = 100μeV with circles at similar locations as a.

c. Measured $g(\boldsymbol{q}, E)$ pattern at E = 100μeV with circles at similar locations as a. The angular width of maxima at $q_j$; $j = 3,4$ in this image indicate that minima in $\Delta_\alpha(\boldsymbol{k}), \Delta_\beta(\boldsymbol{k})$ occur at less than approximately 0.05 radian from the $(0,0) \rightarrow (\pm 1, \pm 1)\pi/a$ $\boldsymbol{k}$-space lines.

d. Superconducting energy-gap $\Delta_i(\boldsymbol{k})$ structure of Sr$_2$RuO$_4$ consistent with the $g(\boldsymbol{q}, E)$ data presented in Figs 4,5.

**Methods:** Supplementary information contains descriptions of the simulations of Bogoliubov quasiparticle interference for $Sr_2RuO_4$, demonstration of preferential tunneling to the $\alpha$- and $\beta$–bands, the experimental techniques for imaging Bogoliubov quasiparticle interference, measurement of the angular distance of gap minima/nodes from $(0,0) \rightarrow (\pm1, \pm1)$ lines , and the analysis of possible energy gap minima elsewhere in $k$-space. The data shall be available upon request to the corresponding author.

**Acknowledgements:** We are grateful to B.M. Andersen, P. Coleman, C. Hicks, B. Ramshaw, S.A. Kivelson, S.H. Simon and A.-M. Tremblay, for very helpful discussions and communications. Y.M. acknowledges support from the JSPS KAKENHI Nos. JP15H05851, JP15K21717, and from the JSPS Core-to-Core Program. V.M. acknowledges funding from the US Department of Energy, Office of Basic Energy Sciences, under Award Number DE-SC0014335. R.S. and A.K. acknowledge support from the US Department of Energy, Office of Basic Energy Sciences, under contract number DEAC02-98CH10886. J.C.S.D and S.D.E. acknowledge support from the Gordon and Betty Moore Foundation's EPiQS Initiative through Grant GBMF4544. J.C.S.D. acknowledges support from Science Foundation Ireland under Award SFI 17/RP/5445, and from the European Research Council (ERC) under Award DLV-788932.

**Author Contributions:** V.M., A.P.M. and J.C.S.D. conceptualized the project. R.S., S.D.E, A.K. and Z.W. carried out the experiments and data analysis. Y.M. and C.S. synthesized the sequence of samples. V.M., A.P.M. and J.C.S.D. supervised the investigation and wrote the paper with key contributions from R.S., S.D.E and Z.W.

**Author Information:** Correspondence and requests for materials should be addressed to to J.C.S.D. at jcseamusdavis@gmail.com. or V.M. at vm1@illinois.edu

Figure 1

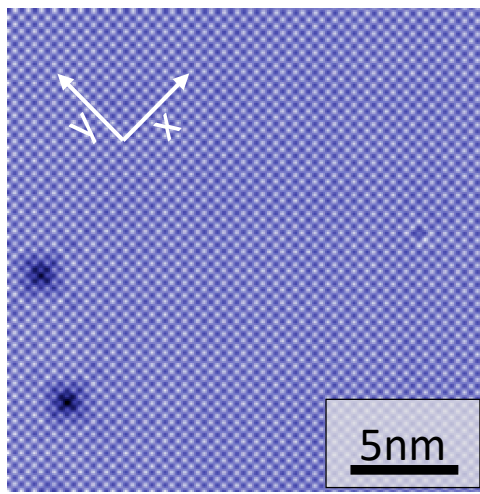
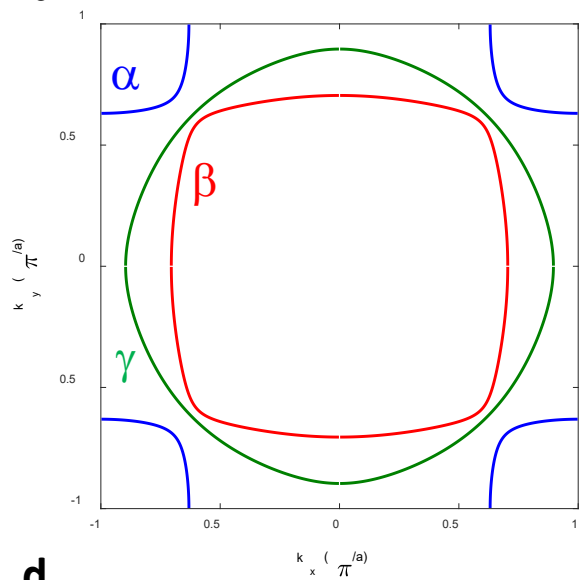
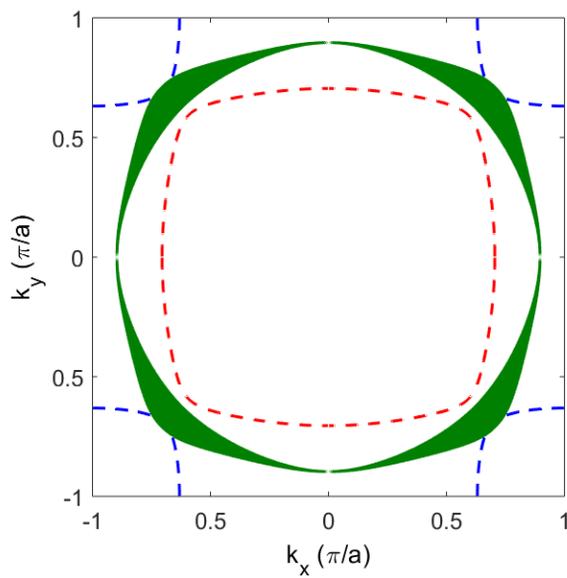
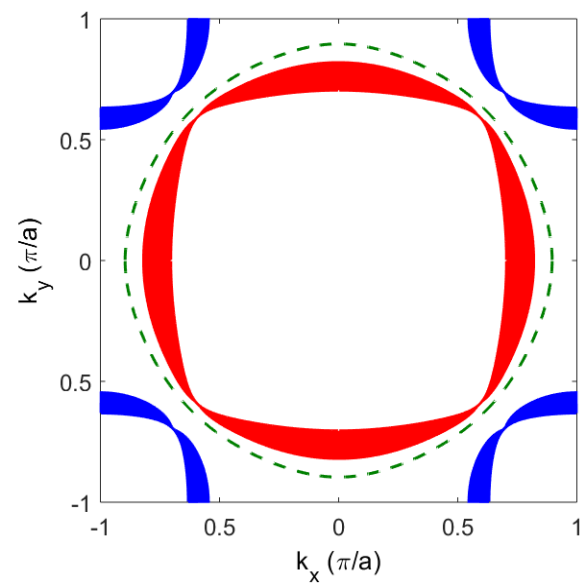

Figure 2

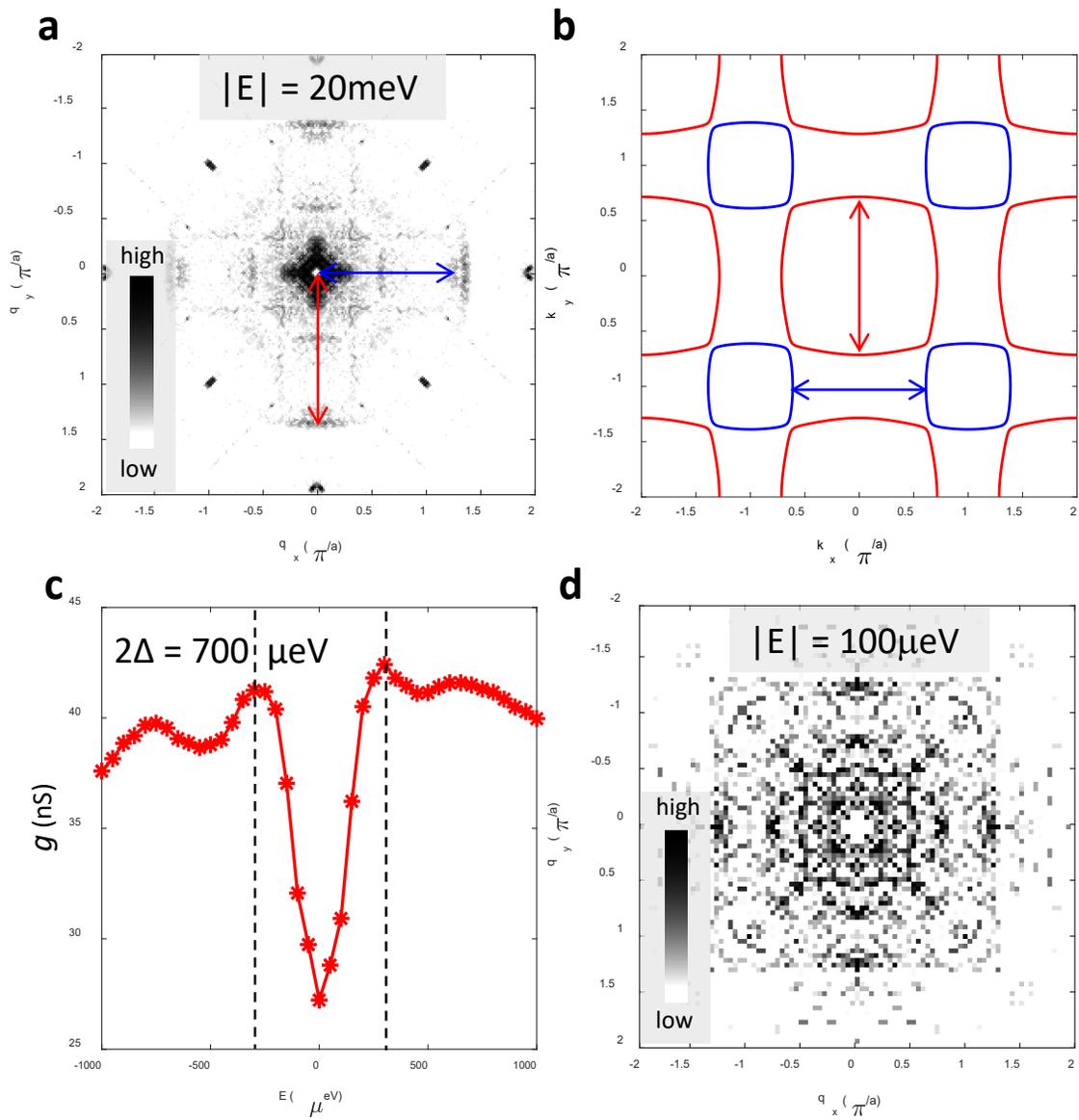

Figure 3

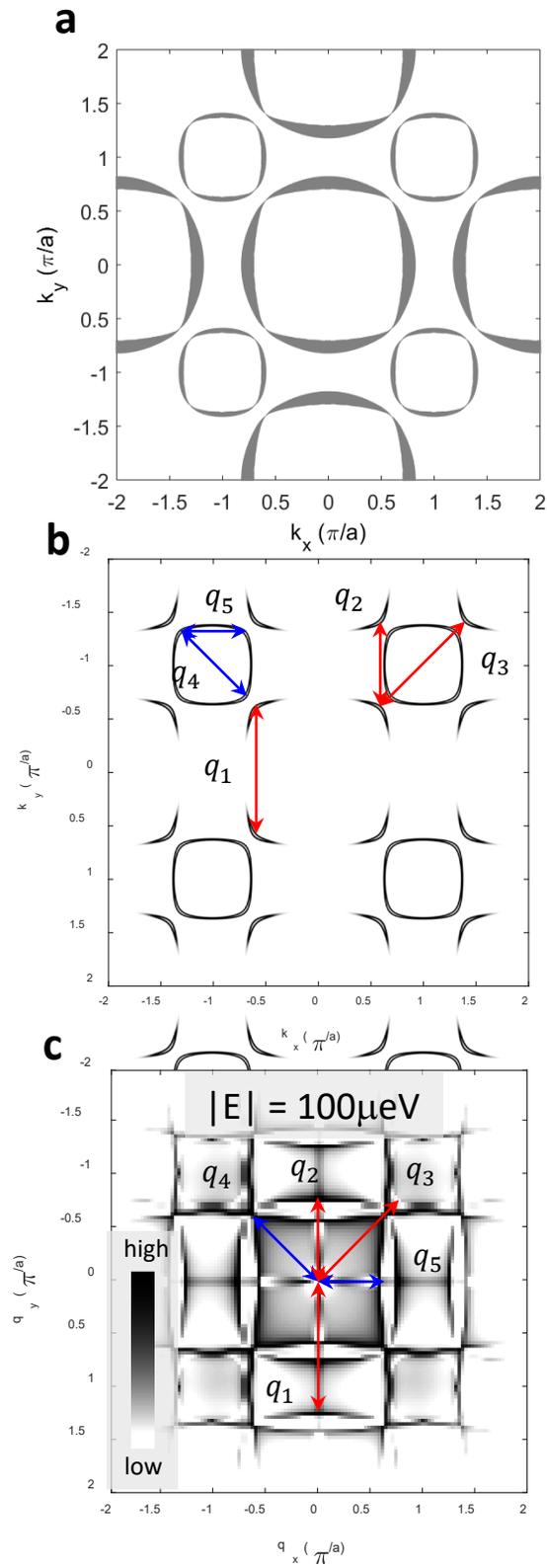

Figure 4

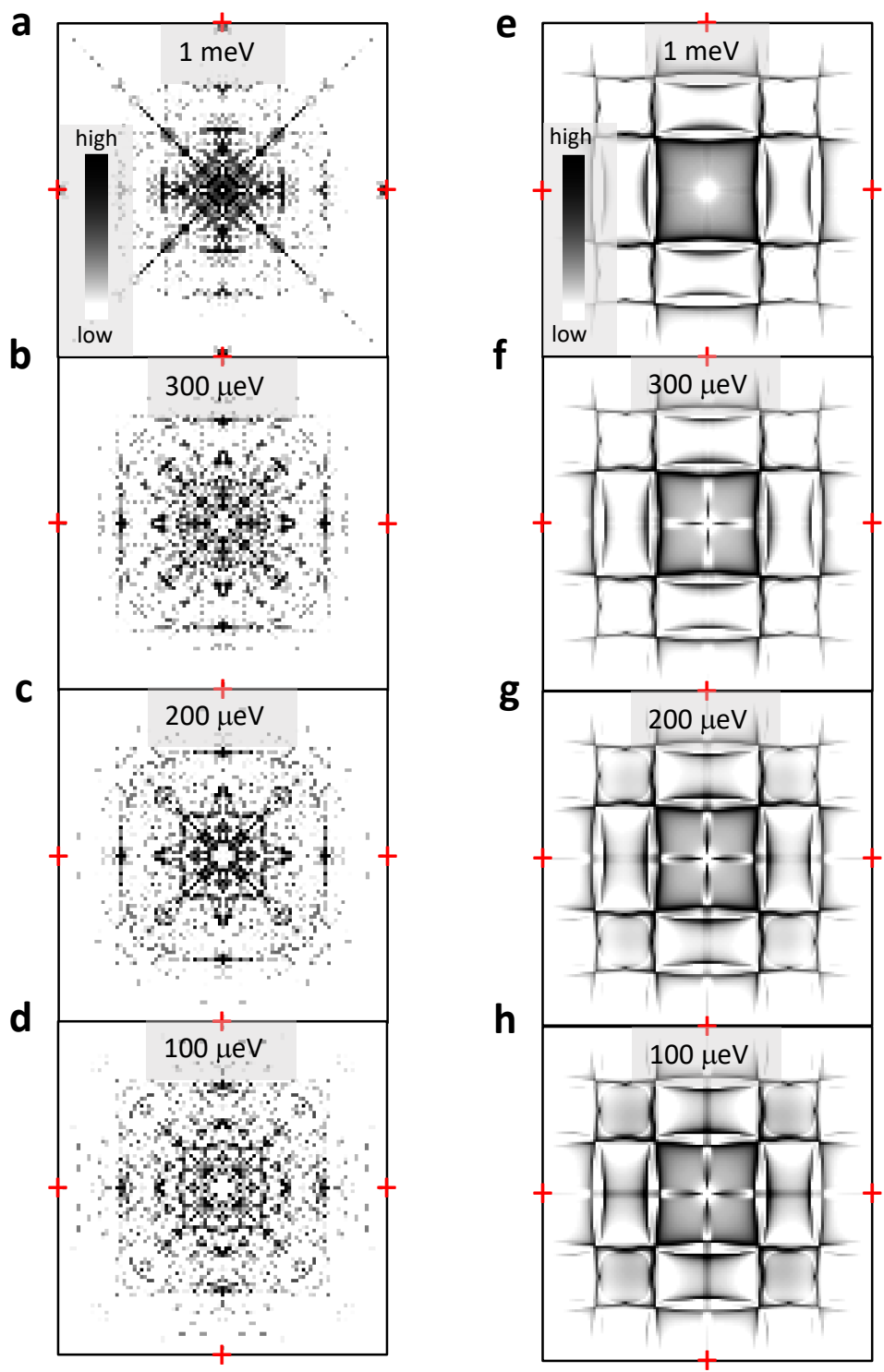

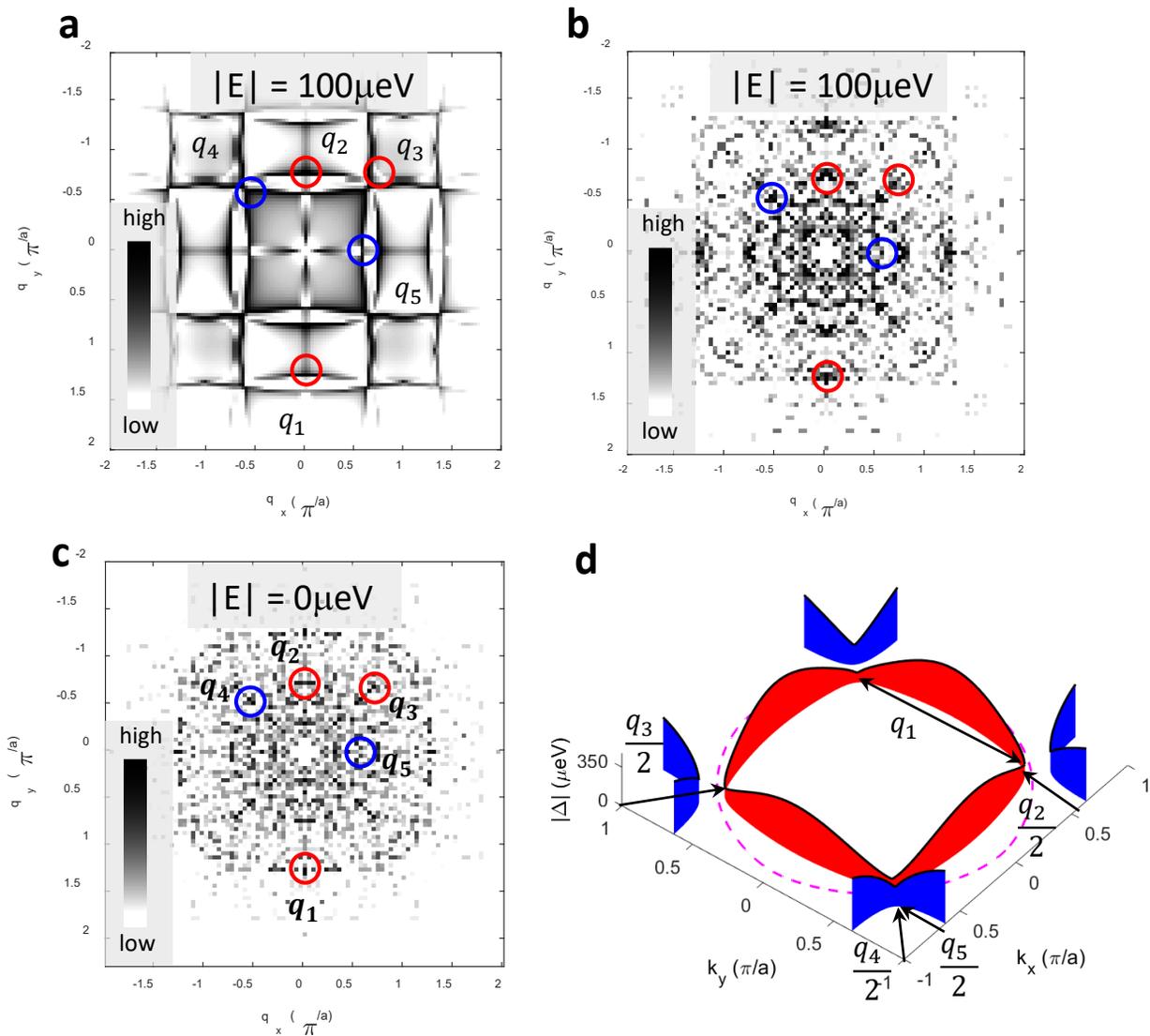

Supplementary Information for

# Momentum Resolved Superconducting Energy Gaps of Sr$_2$RuO$_4$ from Quasiparticle Interference Imaging

Rahul Sharma, Stephen D. Edkins, Zhenyu Wang, Andrey Kostin, Yoshiteru Maeno, Andrew P. Mackenzie, J.C. Séamus Davis and Vidya Madhavan

## I) Simulations of Bogoliubov Quasiparticle Interference for Sr$_2$RuO$_4$

In this section, we describe the multi-band Bogoliubov quasiparticle interference (BQPI) simulation techniques used to create pedagogical multiband $\Delta_i(k)$ models for discussion in the context of Sr$_2$RuO$_4$. Previous researchers have considered the quasi-1D α: β bands and the quasi -2D γ bands separately[1,2,3,4,5,6] and we follow that in our treatment here. We start with a tight binding model.

$$E_{xz} = -\mu_0 - 2t\cos(k_x) - 2t_\perp \cos(k_y) \qquad [S1a]$$

$$E_{yz} = -\mu_0 - 2t\cos(k_y) - 2t_\perp \cos(k_x) \qquad [S1b]$$

$$V_{hyb} = -2V_m \sin(k_x) \sin(k_y) \qquad [S1c]$$

$$\epsilon_\alpha = \frac{1}{2}\left[(E_{xz} + E_{yz}) - \sqrt{(E_{xz} - E_{yz})^2 + 4V_{hyb}^2}\right] \qquad [S1c]$$

$$\epsilon_\beta = \frac{1}{2}\left[(E_{xz} + E_{yz}) + \sqrt{(E_{xz} - E_{yz})^2 + 4V_{hyb}^2}\right] \qquad [S1d]$$

$$\epsilon_\gamma = -\mu_z - 2t_z(\cos(k_x) + \cos(k_y)) - 4t'_z \cos(k_x)\cos(k_y) \qquad [S1e]$$

where we used parameters in units of t $(\mu_0, t_\perp, V_m, \mu_z, t_z, t'_z) = (1.0, 0.1, 0.1, 0.55, 0.2, 0.7)$ and we take $t = 100 meV$ as used in Ref. 7. For clarity, we have treated α:β-bands and γ-band in two separate analyses. The Hamiltonian is given as $H(k) = \sum_k \psi^\dagger(k)\hat{H}(k)\psi(k)$.

For α:β–band model, we choose a basis $\psi^\dagger(k) = (c^\dagger_{\alpha,\mathbf{k}\uparrow}, c_{\alpha,-\mathbf{k}\downarrow}, c^\dagger_{\beta,\mathbf{k}\uparrow}, c_{\beta,-\mathbf{k}\downarrow})$ which leads to the form

$$\widehat{H}_{\alpha:\beta}(k) = \begin{pmatrix} \epsilon_\alpha(k) & \Delta_\alpha(k) & 0 & 0 \\ \Delta_\alpha^*(k) & -\epsilon_\alpha(k) & 0 & 0 \\ 0 & 0 & \epsilon_\beta(k) & \Delta_\beta(k) \\ 0 & 0 & \Delta_\beta^*(k) & -\epsilon_\beta(k) \end{pmatrix} \quad [S2a]$$

For γ-band, we only have a single band which leads to

$$\widehat{H}_\gamma(k) = \begin{pmatrix} \epsilon_\gamma(k) & \Delta_\gamma(k) \\ \Delta_\gamma^*(k) & \epsilon_\gamma(k) \end{pmatrix} \quad [S2b]$$

We used $\Delta_{\alpha:\beta} = \Delta_0 \big(cos(k_x) - cos(k_y)\big)$ to simulate a gap with minima along $(\pm 1, \pm 1)$ directions and $\Delta_\gamma = \Delta_0 \big(sin(k_x) + i sin(k_y)\big)$ to simulate minima along $(1,0); (0,\pm 1)$. This choice of directions of nodes is to compare with the existing models in the literature[1,2,3,4,5,6]. The nodes on the $\alpha:\beta$ bands are proposed along $(\pm 1, \pm 1)$ due to the observed incommensurate antiferromagnet fluctuations at $(0.6\pi, 0.6\pi, 0)$[8].

The unperturbed Green's function is given as $G^0(k, \epsilon) = \big[(\epsilon + i\delta)I - \widehat{H}(k)\big]^{-1}$.

One may consider both intraband and interband scattering for α:β bands and write down the T-matrix for the α:β model as:

$$T_{\alpha:\beta(\omega)}^{-1} = \sigma_z \otimes (V_{intra}I + V_{inter}\sigma_x)^{-1} - \int \frac{dk}{2\pi} G^0(k, \omega) \quad [S3a]$$

Where $\sigma_i$ denote the Pauli matrices. We take $(V_{intra}, V_{inter}) = (1.0, 1.0)$ in units of t. The T-matrix for γ model is given as:

$$T_\gamma^{-1}(\omega) = (V_{intra}\sigma_z)^{-1} - \int \frac{dk}{2\pi} G^0(k, \omega) \quad [S3b]$$

The scattering problem for a single impurity at the origin can be solved in first order Born approximation[9,10] to calculate the change in density of states as:

$$\delta N(q, E) = -Im\left[Tr\left(\int \frac{dk}{2\pi} G^0(k+q, \omega) T(\omega) G^0(k, \omega)\right)\right] \quad [S4]$$

The resolution of the momentum-space grid used is critical in the numerical evaluation of $\delta N(q, E)$. One needs to achieve $\delta k \ll \pi v_F/v_\Delta$ to be able to accurately capture the contribution of constant contours of energy (CCE) resulting from in-gap Bogoliubov quasiparticle states. For our calculations, we used a grid of 8000x8000 pixels for $k \in (\pi/a, -\pi/a)$. We apply repeated zone scheme to our $\delta N(q, E)$ image (Fig. S1a) to create a bigger reciprocal space which results in q-space of $q \in (-2\pi/a, 2\pi/a)$ (Fig. S1b). We then apply a structure factor $S(q) = 2\sqrt{[1 + cos(q_x/2)[1 + cos(q_y/2)]]}$ which reflects fourfold rotationally symmetric electronic structure[11], whose high-q features cannot be detected by a finite size tip (Fig. S1c). Finally, we repeatedly apply 4x4 pixel averaging to reduce the resulting 16000x16000 pixels image to produce the final image (Fig. S1d).

Fig. S2a-d and S2e-h show images generated using this scheme for γ-band and α:β bands respectively, in superconducting state of Sr$_2$RuO$_4$. The clear differences in the shape of $\delta N(\boldsymbol{q}, E)$ can be observed. The $\delta N(\boldsymbol{q}, E)$ for the γ-band in Fig. S2 a-d has shape arising from the scattering within the underlying normal state quasi-circular γ-band which is in clear contrast to $\delta N(\boldsymbol{q}, E)$ for α:β bands in Fig. S2e-h arising from scattering within the quasi-square shaped α:β bands. The different shape and size of γ-band and α:β bands lead to strong scattering intensity at very different locations in $q$-space as can be seen from an overlay of the strong scattering features $q_j, j = 1..5$ in the $\delta N(\boldsymbol{q}, E)$ from α:β bands in Fig. S2h onto $\delta N(\boldsymbol{q}, E)$ from γ-band in Fig. S2d. There is no observed strong intensity feature at any of $q_1, q_2, q_4$ and $q_5$ in Fig. S2d, meaning that the γ-band BQPI is not detected by our measurements. Nevertheless, the $\delta N(\boldsymbol{q}, E)$ we simulate are in good agreement with previous theoretical calculations in Ref.1 Although in Ref. 1, all three bands were considered simultaneously, the features from α:β bands as shown in Fig. 5 of Ref. 1 are quite consistent with our simulations in Fig. S2 f-h.

## II) Preferential tunneling to the α- and β-bands of Sr$_2$RuO$_4$

In this section, we describe our analysis of quasiparticle interference data recorded at T=2.1K in a 256x256 grid of pixels from E=-20meV to E=+20meV with junction setup at V=20meV and I=40pA. The integrated density of states at a given height of tip at setup bias and current, leads to the so called "setup effect"[12,13] which strongly affects the QPI. To overcome the setup effect, we perform a per-pixel division of the measured $g(\boldsymbol{r}, E = eV)$ map at each E by the current map I($\boldsymbol{r}$,E=eVs) measured at setup voltage Vs. The resulting setup corrected image shows vivid QPI as shown in Fig. S3a. We take Fourier transform of these images to discover the scattering wavevectors. Fig. S3b shows such a representative layer at E=20meV. We identify the scattering wavevectors here which we used in our analysis as intraband scattering in β-band (shown in red) and interband scattering in α-band (shown in blue). The other q-peak denoted with an orange arrow is the interband beta scattering. The rest of the features in Fig. S3b either do not disperse or do not appear at all energies hence are not considered in the analysis. The arrows showing these experimentally observed wavevectors are placed on the Fermi surface in Fig. S3c. This enables us to extract **k**-vectors from **q**-vectors using following relations:

$$k_\alpha = \frac{\pi}{a} - \frac{q_\alpha}{2} \qquad [S5a]$$

$$k_\beta = \frac{q_\beta}{2} \qquad [S5b]$$

We extract the *q*-points for these wavevectors as shown in Fig. S2c at each energy layer, get the k-values using eq. S5 and fit a tight binding model as described

To compare our results with other experiments, in Fig. S2d we overlay the ($k_x$,$k_y$) points extracted from the E=0 layer in our normal-state QPI experiments on the ARPES Fermi surface[14], showing a match to the α– and β-bands within the error bars; any contributions from the surface states[15] due to termination of these bands does not alter this identification. Moreover, in the table below, we compare our results with quantum oscillation experiments[16]. For such comparison, it should be kept in mind that our analysis uses *q*-vectors which are observed with maximum intensity along [1,0] direction, while quantum oscillation provides quantities averaged over the whole Fermi Surface.

| Quantity | Our Experiment | Mackenzie et al.[16] |
|---|---|---|
| $k_{F,\alpha}$ measured at $E = 0$ (1/A) | $0.30 \pm 0.02$ | $0.30 \pm 0.0006$ |
| $k_{F,\beta}$ measured at $E = 0$ (1/A) | $0.55 \pm 0.02$ | $0.621 \pm 0.008$ |

**Table S1.** Comparison of averaged $k_F$ over whole FS calculated using tight binding fit to experimental data and averaged $k_F$ measured in quantum oscillations experiments.

In combination, these observations show that during our studies in which the $Sr_2RuO_4$ crystal was terminated by the SrO layer only (Fig. 1a), tunneling occurs preferentially to the $α$- and $β$–bands, and that these can be distinguished from each other in experimental data. We note that the extended features as seen in Fig. S3b are very different from the features deep within superconducting gap as shown in Fig. S5, where sharp spots are seen.

## III) Imaging Bogoliubov Quasiparticle Interference

To study the Bogoliubov quasiparticle interference (BQPI) in superconducting state, we use our SI-STM at T=90mK to record $g(r,E) = dI(r, E = eV)/dV$ at each pixel $r$ for multiple energies $E$ to generate a real space map of LDOS $g(\boldsymbol{r}, E)$. The mixing chamber of dilution fridge is thermalized with the STM head using a custom built thermal short (electrically isolating) which has proved reliable for previous heavy-fermion[17] and SJTM[18] studies. The Cernox thermometer is mounted on the sample stage itself, within about 2 mm of close the sample stud. These careful thermalization and thermometry steps ensure that the sample crystal is indeed measured at 90mK.

We operate in standard constant current mode and at each pixel $r$, we adjust the height with a feedback loop to reach a constant current $I(r, V_{setup})$ at a setup voltage $V_{setup}$. Then we turn the feedback off and measure dI/dV spectrum using standard Lock-In techniques by applying a small bias modulation. Figure S4a shows the topography recorded simultaneously while recording the real space LDOS map $g(\boldsymbol{r}, E)$ for each pixel. Two impurity atoms in our field of view are circle are highlighted by dashed red circles. Due to interplay of tip height and integration of density of states up to setup voltage, the well-documented *setup effect*[12] is unavoidable. In weakly dispersing system, like $Sr_2RuO_4$ close to Fermi surface, the setup voltage and current affects the BQPI patterns very significantly. There are multiple schemes to counter this setup effect[19]. We employ a *setup-correction* and divide our $g(\boldsymbol{r}, E)$ data by the $I(\boldsymbol{r}, V_{setup})$ to reveal Bogoliubov quasiparticle interference.. Figure S4b shows these setup-corrected $g(\boldsymbol{r}, E)$ images using an absolute intensity scale, showing how the signal diminishes inside the energy gap. Figure S4c shows measured $g(\boldsymbol{r}, E)/I(\boldsymbol{r}, V_{setup})$ using a scale self-normalized for each image, making clearer the spatial variation of tunnel conductance at every energy.

Fig. S5 presents the Fourier transform of the setup corrected $g(r, E)$ images. We present both unsymmetrized and symmetrized images on a linear colorscale here without cutting off any intensity. The colorscale used for Fig. S5 is shown in Fig. S6b while Fig. S6a shows the colorscale used in all the other BQPI figures in SI and main text. Fig. S5c and S5g contain circles to guide the eye for the features $q_j : j = 1,2,..5$ as defined in the main text. From Fig. S5a-d, it can be seen that the features $q_j : j = 1,2,..5$ really do exist in the data. and are not resulting from symmetrization or tweaking intensity. Symmetrization and intensity cutoff for contrast adjustment is performed to enhance the clarity of already existing peaks in the BQPI data to produce Fig. 4a-d in the main text.

## IV) Angular distance of gap minima/nodes from $(0, 0) \rightarrow (\pm 1, \pm 1)$ Lines

The width of $g(q, E)$ features can be analyzed to put an upper limit on how far the deduced minima/nodes are from the $k$-space symmetry lines along $(0,0) \rightarrow (\pm 1, \pm 1)\pi/a$. In Fig. S7a we indicate the scattering along $(\pm 1, \pm 1)$ which we observe in our experiment as features $q_3$ and $q_4$. As can be seen, if there are minima/nodes on the Fermi surface, the scattering would be subtended by angle 2θ about the X-point. This angle can be estimated from the width $w$ of the feature and the length of the scattering vector $d$ as shown for $q_3$ in Fig. S7a. In Fig. S7b we show the measured $g(q, 0)$ features $q_3$ and $q_4$, and in Fig. S7c (S7d) the zoomed versions of $q_3(q_4)$. The angle subtended about the X-point is then determined by $2\theta = atan(w/d)$, where $w$ is the width of the feature as shown in Fig. S7c and S7d and $d$ is the distance from the center of the $q$-space to the central pixel in the feature. As shown in Fig. S7c and S7d we find that $2\theta_{q_3} = 0.18$ rad and $2\theta_{q_4} = 0.22$ rad. Therefore, when considering angles about the lines $(0,0) \rightarrow (\pm 1, \pm 1)\pi/a$ but measured from the Γ-point, these same minima/nodes on α:β bands subtend a maximum angle $2\theta_\Gamma$ of approximately 0.1 radians. Thus, all eight gap minima/nodes have a maximum angular displacement $\theta$ measured about (0,0), of within approximately $\pm 0.05\ rad$ away from the $(0,0) \rightarrow (\pm 1, \pm 1)\pi/a$ lines

## V) Energy gap minima elsewhere in $k$-space.

In this section, we compare all features in our $g(\mathbf{q}, E)$ data deep within the SC gap ($E = 100\mu eV$) to the complete simulation for the energy-gaps $\Delta_\alpha(\mathbf{k})$ and $\Delta_\beta(\mathbf{k})$ shown in main-text Fig. 5d. In Fig. S8a we show an overlay of $g(\mathbf{q}, 100\mu eV)$ data on the simulation $\delta N(\mathbf{q}, 100\mu eV)$ for the energy-gaps as shown in Fig. 5c. The low-$q$ area $0 < |q_x, q_x| < \pi/2a$ is not considered because (as typical in SISTM studies) these regions are dominated by long range disorder/scan-drift. Comparison (for the region $\pi/2a < |q_x, q_x| < 3\pi/2a$) of the features in $g(\mathbf{q}, 100\mu eV)$ and in $\delta N(q, 100\mu eV)$ in the format shown in Fig. S8a yields very good visual agreement. This indicates that the structure of $\Delta_\alpha(\mathbf{k})$ and $\Delta_\beta(\mathbf{k})$ shown in Fig. 5d of the main text is sufficient to explain virtually all the observed features of the in-gap $g(\mathbf{q}, E)$ data, and no other deep gap minima elsewhere in $\mathbf{k}$-space are required.

# SI Figure Captions

### Fig. S1 Modeling the BQPI of $Sr_2RuO_4$

a. Image calculated using Eqn. S4 for the pedagogical model as described in SI section I at E=100 µeV.
b. Image generated after applying repeated zone scheme to a.
c. Image after applying a structure factor as described in SI section I.
d. Final image generated by repeatedly applying 4 pixel averaging to get 125x125 pixel image (as in the $g(r,E)$ measurements) from 16000x16000 pixel image

### Fig. S2 BQPI simulations for γ and α:β bands

a.-d. Predicted $\delta N(q,E)$ images generated as described in SI section I for γ-band for E = 1meV, 300µeV, 200µeV and 100µeV. Red crosses denote Bragg peaks.

e.-h. Predicted $\delta N(q,E)$ images generated as described in SI section I for α:β-bands for E = 1meV, 300µeV, 200µeV and 100µeV. Red crosses denote Bragg peaks.

### Fig. S3 Analysis of Normal State of $Sr_2RuO_4$

a. Vivid QPI oscillations as seen in setup corrected g(r,E=20meV) image at recorded at T=2.1K as described in SI section II.
b. Fourier Transform of image in a. Arrows indicate the major scattering features. Red (Blue) arrow indicates intra (inter) beta (alpha) band scattering as shown in c. Orange arrow indicates inter-beta band scattering.
c. Fermi Surface of $Sr_2RuO_4$ with major scattering vectors as identified in b. overlaid.
d. The $(k_x, k_y)$ points for beta (pink) and alpha (cyan) bands calculated using eq. S5 from the q-vectors identified from E=0 layer overlaid on ARPES Fermi Surface[14].

**Fig. S4 Real Space Bogoliubov Quasiparticle Interference**

a. The topograph recorded at T=90mK simultaneously with the $g(r,E) = dI/dV(r,E)$ measurements. Red dashed circles denote the impurity atoms.

b. The setup corrected $g(r,E)$ images. Spectroscopic setup conditions were I=40pA and V=1mV.

c. The images presented in b. with a self-normalized color scale for each image

**Fig. S5 Real Space Bogoliubov Quasiparticle Interference**

a.-d. Unsymmetrized Fourier transform of setup corrected $g(r,E)$ with linear colorscale and no intensity cutoff.

e.-h. Symmetrized Fourier transform of setup corrected $g(r,E)$ from a.-d. wiith linear colorscale and no intensity cutoff.

**Fig. S6 Colorscale for BQPI figures**

a. Colorscale employed for the BQPI figures presented in this manuscript except S5.

b. Colorscale employed for BQPI figures in S5.

**Fig. S7 Angular Distance of Gap Minima/Nodes from $(0,0) \rightarrow (\pm 1, \pm 1)$ Lines**

a. Schematic of scattering in *k*-space which leads to scattering features $q_3$ and $q_4$ in BQPI.

b. Experimental g(q,E) measured at 0 µeV showing $q_3$ and $q_4$.

c. Zoomed view of $q_3$ showing w and d which were used to determine θ.

d. Zoomed view of $q_4$ showing w and d which were used to determine θ.

**Fig. S8 Detailed Comparison of experimental BQPI pattern and Simulation**

a. Overlay of the experimental g(q,E) measured at 100 µeV and the simulation with a $d_{x^2-y^2}$ order parameter.

Figure S1

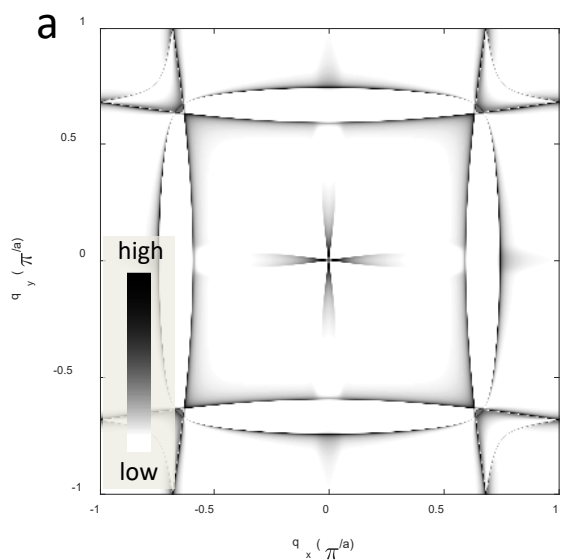 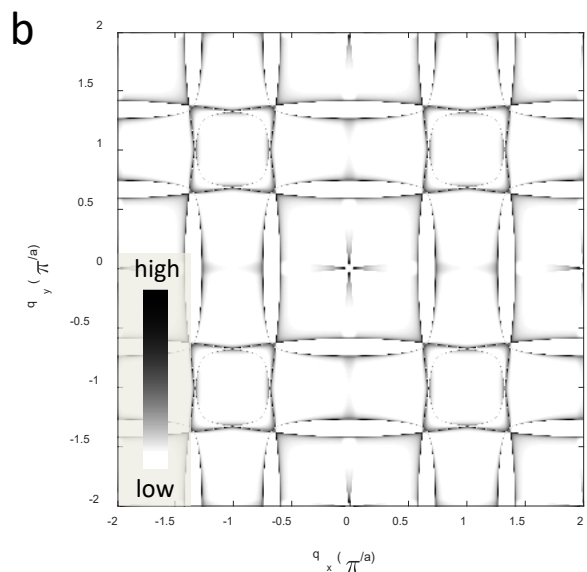
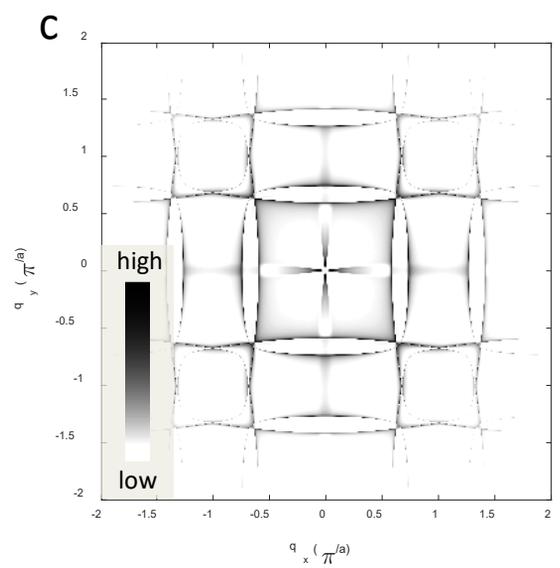 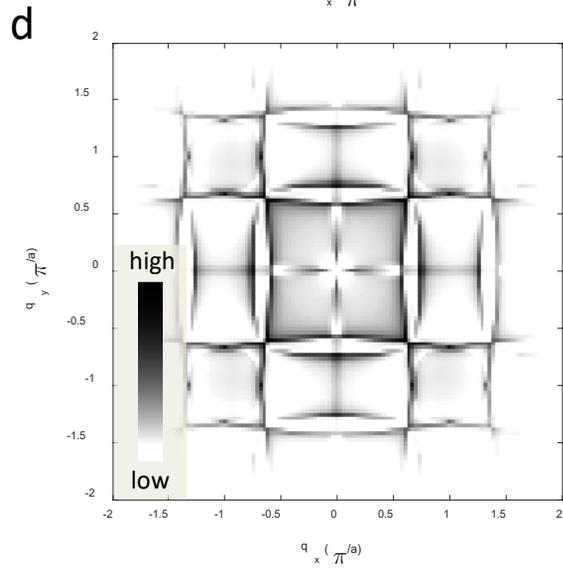

Figure S2

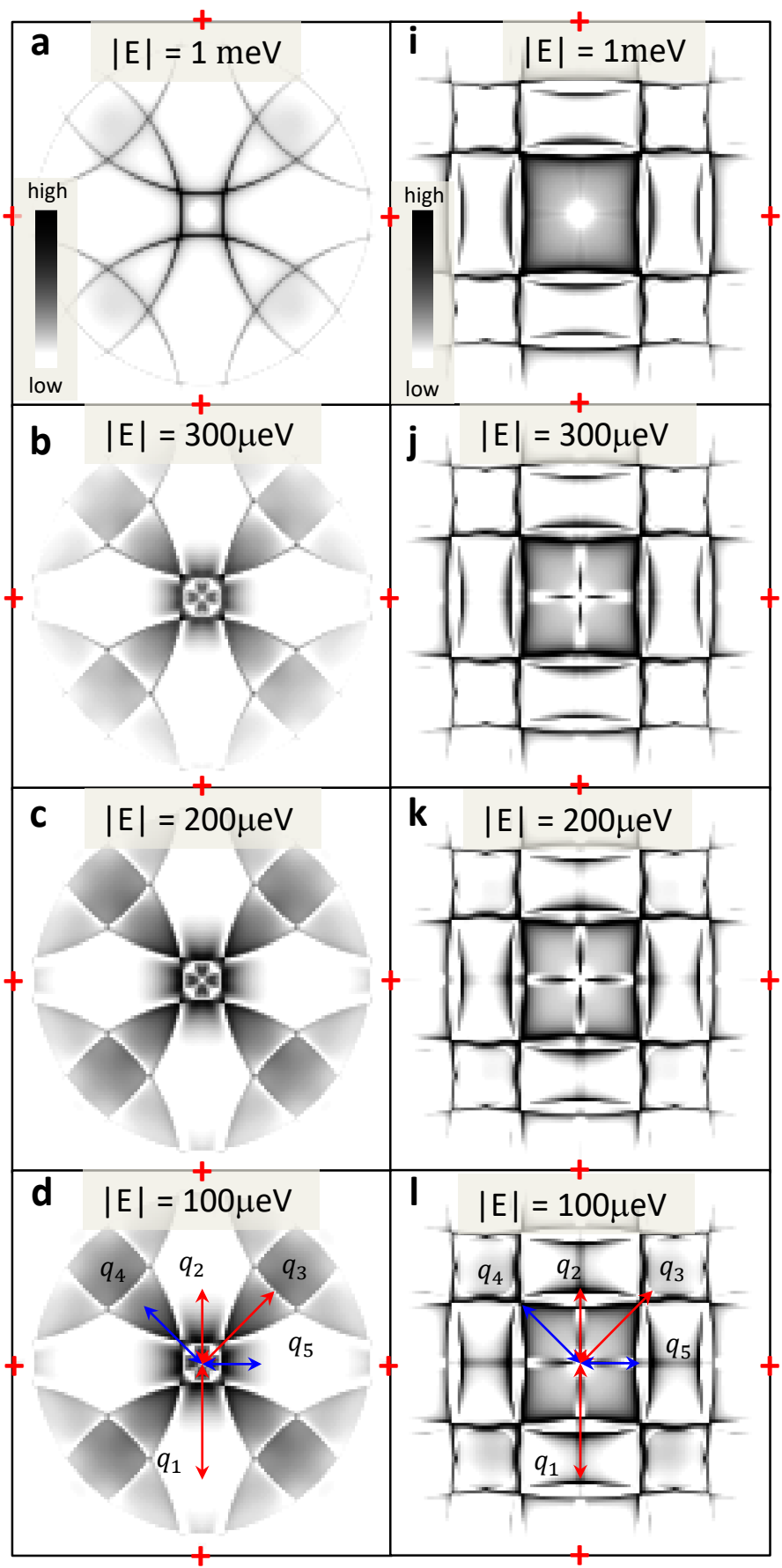

Figure S3

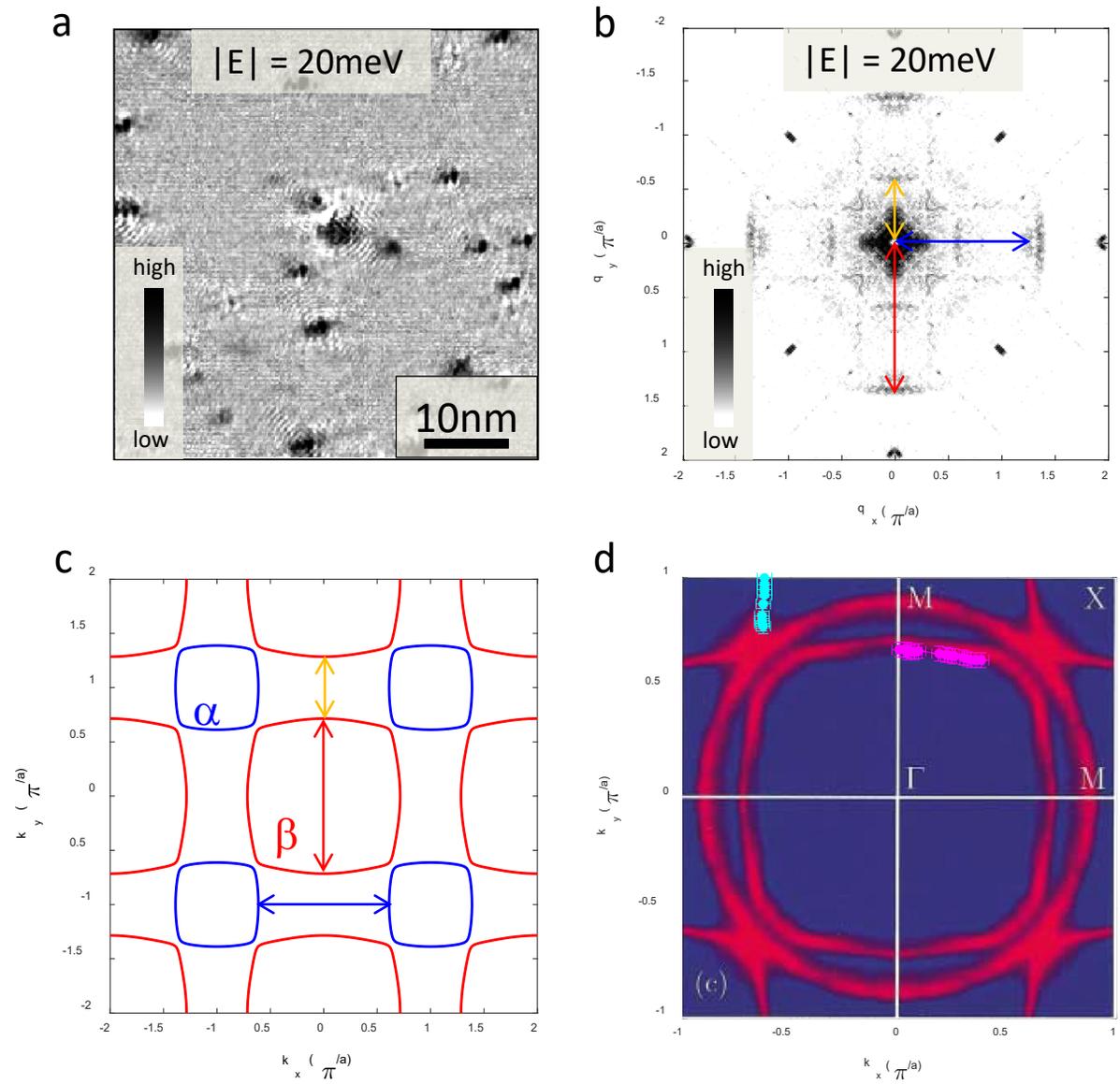

# Figure S4

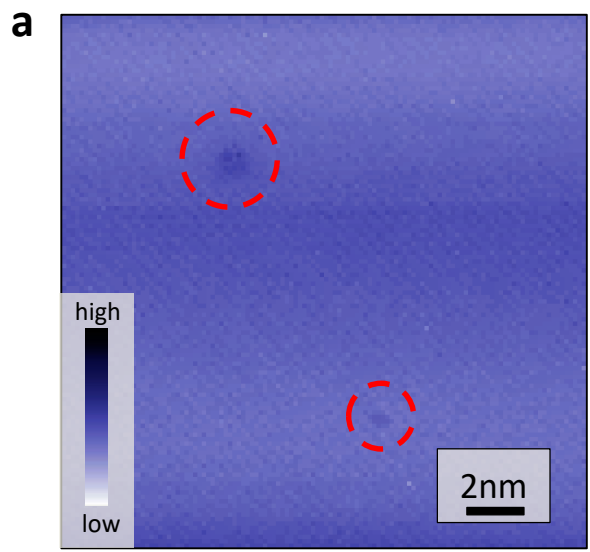

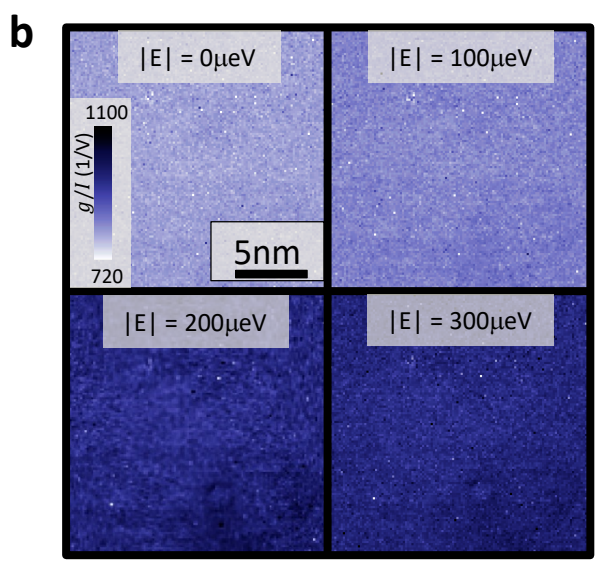

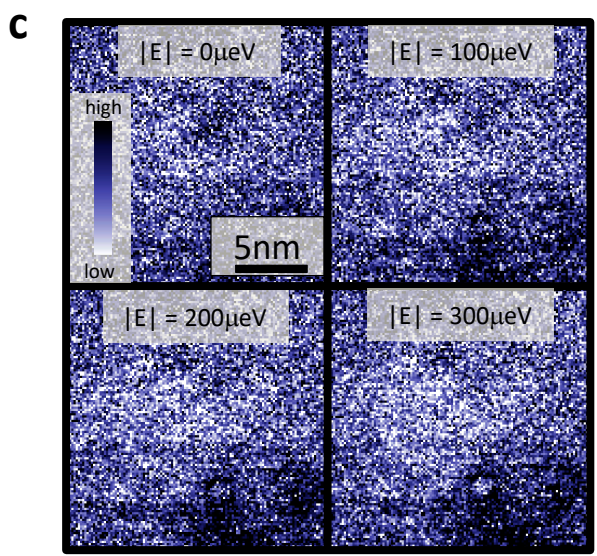

Figure S5

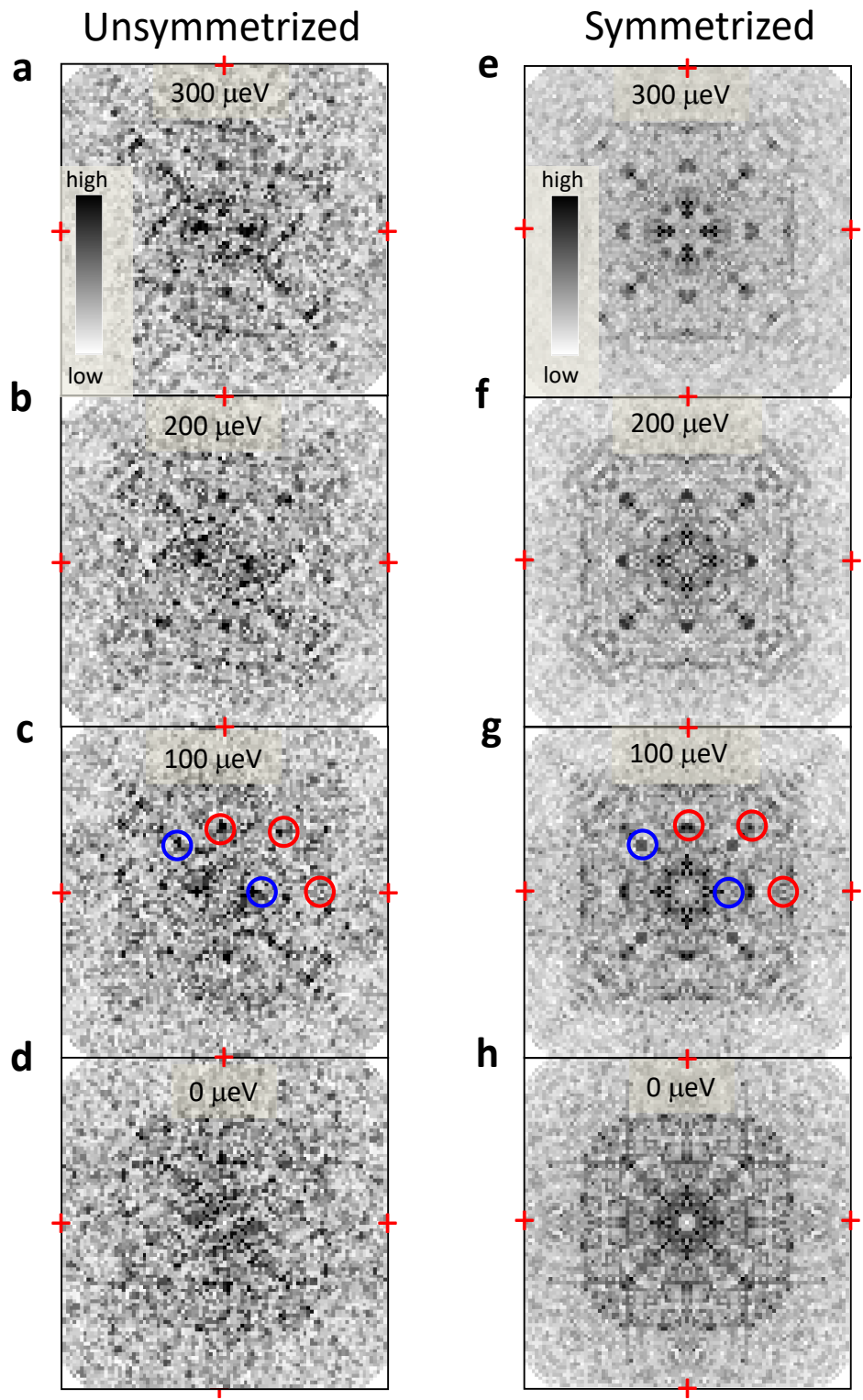

Figure S6

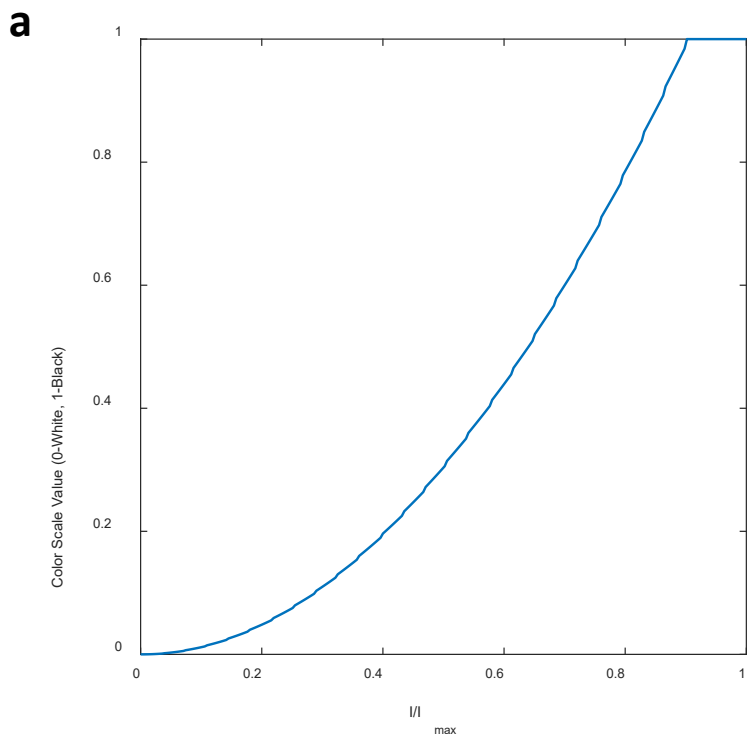

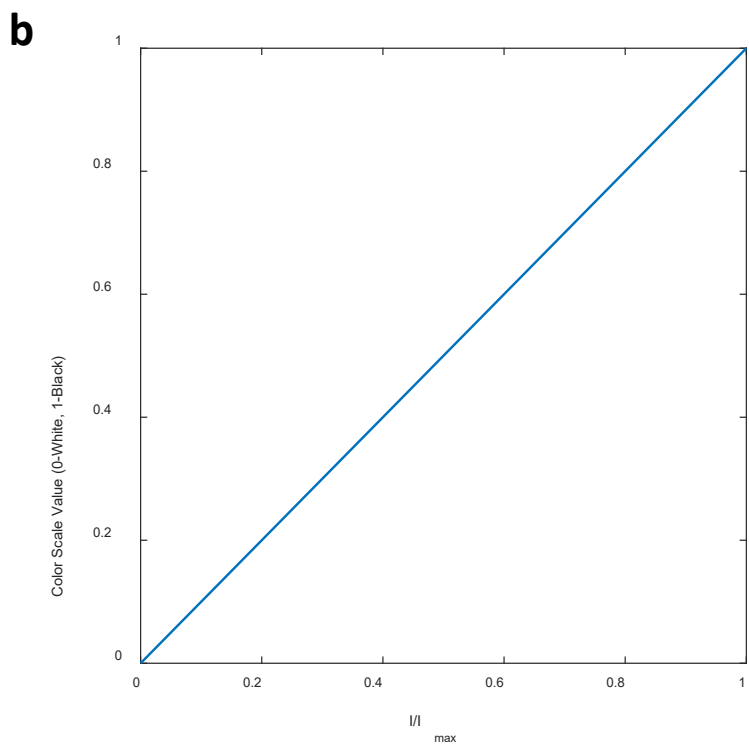

Figure S7

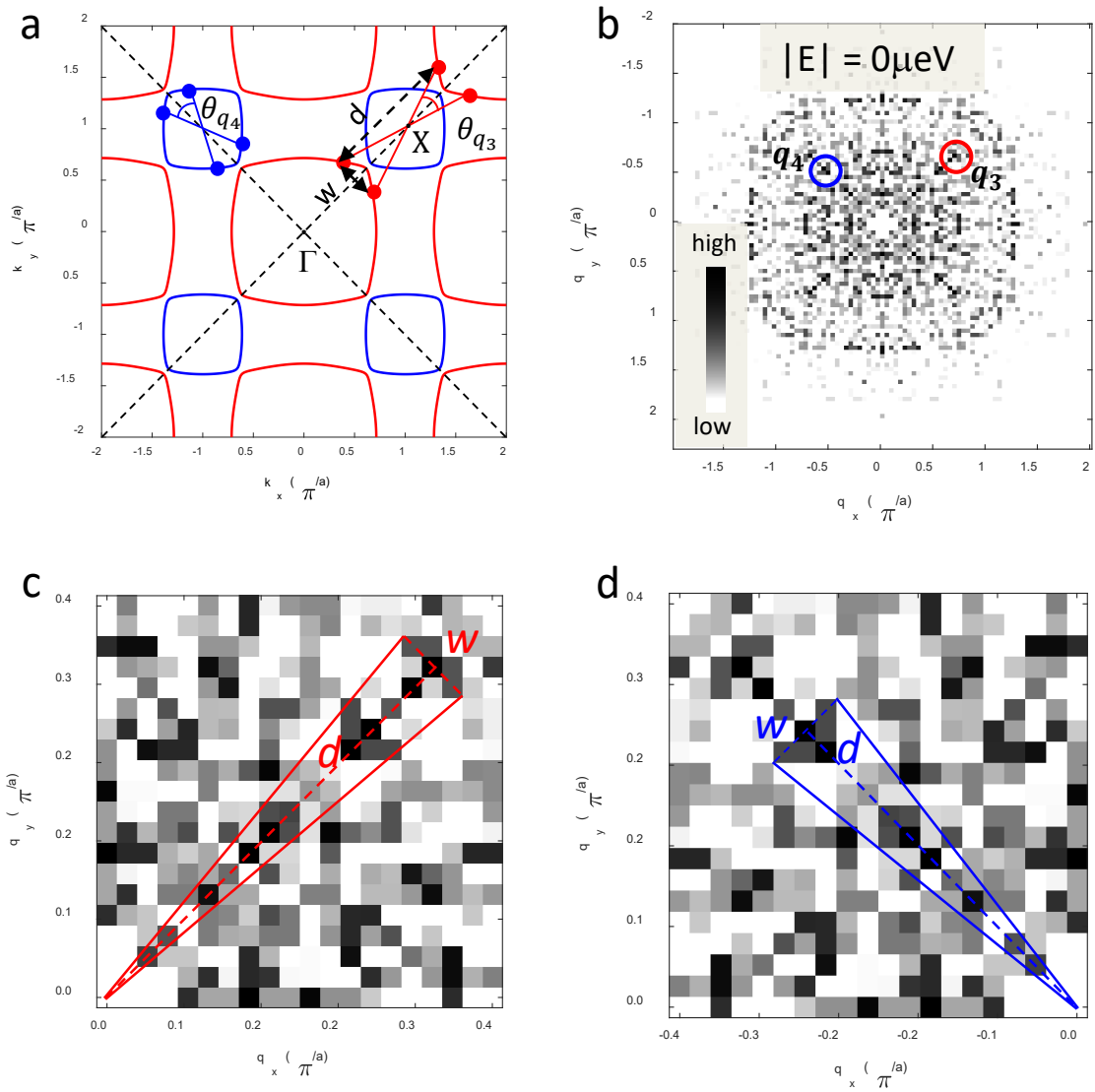

Figure S8

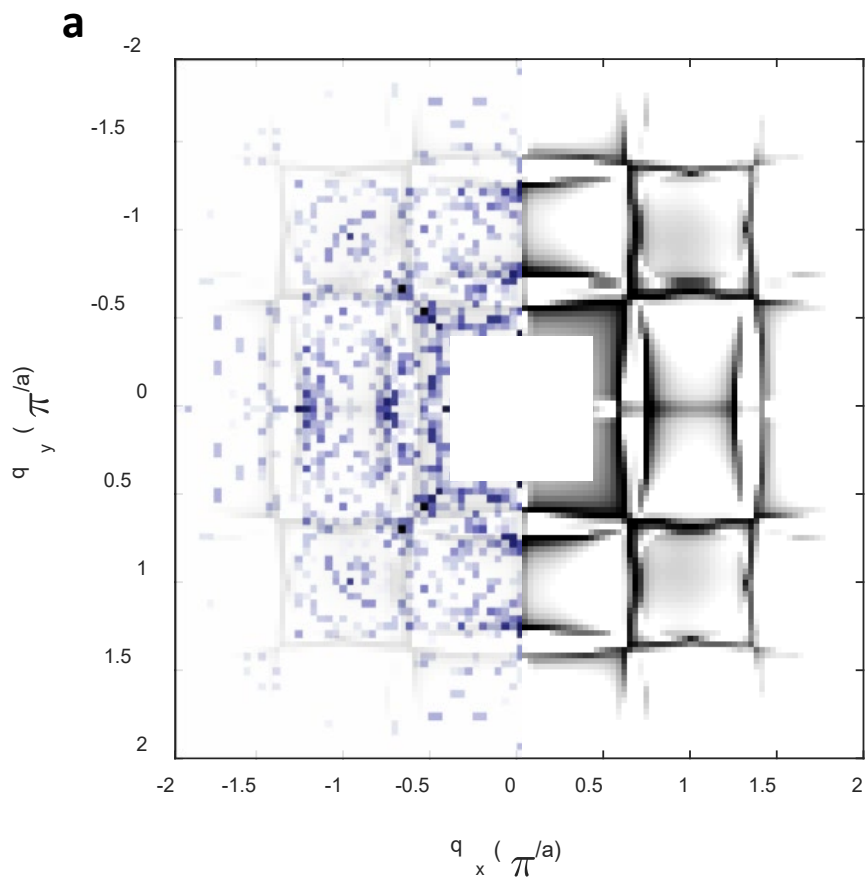